\documentclass[pra, twocolumn, showpacs,
superscriptaddress , unsortedaddress]{revtex4}

\usepackage{latexsym}
\usepackage{amssymb}
\usepackage{exscale}
\usepackage{bm, graphics}
\usepackage{psfrag, bm}
\usepackage{graphicx}

\newcommand{\D}{\mbox{\rm d}}

\def\ket{\rangle}

\begin{document}

\author{M. Khanbekyan}
\email[E-mail address: ]{mkh@tpi.uni-jena.de}
\affiliation{Theoretisch-Physikalisches Institut,
Friedrich-Schiller-Universit\"{a}t Jena, Max-Wien-Platz 1, D-07743 
Jena, Germany}

\author{ L. Kn\"oll}
\affiliation{Theoretisch-Physikalisches Institut,
Friedrich-Schiller-Universit\"{a}t Jena, Max-Wien-Platz 1, D-07743 
Jena, Germany}

\author{A. A. Semenov}
\affiliation{ Fachbereich Physik, Universit\"{a}t Rostock, 
Universit\"{a}tsplatz 3, D-18051 Rostock, Germany}
\affiliation{Institute of Physics, National Academy of Sciences of
Ukraine, 46 Prospect Nauky, UA-03028 Kiev, Ukraine}

\author{W. Vogel}
\affiliation{ Fachbereich Physik, Universit\"{a}t Rostock, 
Universit\"{a}tsplatz 3, D-18051 Rostock, Germany}

\author{D.-G. Welsch}
\affiliation{Theoretisch-Physikalisches Institut,
Friedrich-Schiller-Universit\"{a}t Jena, Max-Wien-Platz 1, D-07743 
Jena, Germany}

\title{
Quantum-state extraction from high-$\boldsymbol{Q}$ cavities 
}
\date{\today}

\begin{abstract} 
The problem of extraction of a single-mode quantum state from a
high-$Q$ cavity is studied for the case in which the time of
preparation of the quantum state of the cavity mode is
short compared with its decay time. The temporal evolution
of the quantum state of the field escaping from the cavity is
calculated in terms of phase-space functions.
A general condition is derived under which the quantum state of the pulse
built up outside the cavity is a nearly perfect copy of
the quantum state the cavity field was initially prepared in.  
The results show that unwanted losses prevent the realization of 
a nearly perfect extraction of nonclassical quantum states
from high-$Q$ optical microcavities with presently available technology.
\end{abstract}


\maketitle

\section{Introduction}
\label{sec1}

High-$Q$ cavity QED has offered a number of novel possibilities
of quantum-state engineering (see, e.g, Refs.~\cite{Raimond01, Pinkse00,
Hood00, Doherty00} and references therein).
In particular, it provides promising tools to generate
nonclassical quantum states of atoms and light for further use. 
Accordingly various applications have been proposed, with
special emphasis on quantum communication and computation
\cite{Knill01, Enk98, Tregenna02}. 
In contrast to atomic states, the usage of quantum states
of light as carriers of quantum information is especially
appropriate, due to the reliability of light to propagate over
long distances \cite{Pan03, Bennett00}. 

Various schemes for generating nonclassical light in
cavities have been considered. They are typically based
on effective two-level atoms. A generator of
single photon Fock states in an active microcavity 
with pump self-regularization has been presented \cite{Martini96}. 
It has also been shown experimentally that an entangled state 
of two nondegenerate cavity modes can be produced with
means of a sequence of differently tuned interactions of a
pair of single atoms with the two cavity modes \cite{Rauschenbeutel01},
and a scheme for entangling two modes of spatially separated cavities
by consecutively passing through them atoms
has been purposed \cite{Browne03}. A scheme has been
proposed \cite{Domokos98} and
experimentally realized \cite{Brattke01}
for the generation of photon number states on demand, 
by subjecting single two-level atoms passing through
a cavity to $\pi$ pulse interaction.
Similarly, a proposal has been made for entangling
two cavity modes via interaction with 
a bunch of two-level atoms assisted by a strong classical
driving field \cite{Solano03}.
Schemes that exploit multilevel atoms
have also been studied both theoretically \cite{Parkins95, Lange00}
and experimentally \cite{Hennrich00},
with special emphasis on effective three-level atoms
of $\Lambda$ type.

The main obstacles
to generate nonclassical states are
the various decoherence effects associated with, e.g.,   
the motion and spontaneous emission of the atoms 
as well as scattering, absorption, and transmission 
of the photon field.
Several proposals have been made to reduce the effect
of decoherence due to the atomic motion \cite{Duan03}. 
In order to reduce unwanted spontaneous emission in schemes
that exploit multi-level atoms, adiabatic transfer techniques
have been promising \cite{Parkins95,Hennrich00,Lange00}. 
Moreover, the adiabatic passage is the main idea of 
the proposal of quantum networks of trapped atoms, 
where cavity modes provide communication channels, 
by leaking out of the cavities and 
propagating via optical interconnectors \cite{Cirac97}.
Scattering and absorption losses, which
are unavoidably connected with any material system,
may be reduced by well designed cavities and the use of
materials showing extremely weak absorption, and
almost perfectly reflecting mirrors reduce the transmission
losses. In particular, in Ref.~\cite{Zippilli03} a method
is presented for the protection of a generic quantum state
of a cavity mode against the decohering effects of photon
losses by feedback atoms crossing the cavity mode. 

On the other hand, transmission losses are necessarily
required to be taken into account
in order to implement high-$Q$ cavities that can serve
as sources that emit nonclassical radiation 
for further use outside the cavities 
\cite{Hennrich00, Kuhn02, Rempe92, Hood01, Pelton02}.     
The natural question arises whether or not nonclassical
states of light, once generated inside a high-$Q$ cavity,
can be extracted from the cavity and what the ultimate limits are.
In the schemes considered, it is often made the
ad hoc assumption of nearly perfect extraction
(see, e.g., Refs.~\cite{Martini96, Lange00, Saavedra00, Cirac97}).
The fact however is that even very small material absorption may be
expected to lead to drastic quantum state degradation \cite{Scheel01}.   

Recently, homodyne detection
of the quantum state of the field leaving a high-$Q$ cavity
has been studied theoretically \cite{Santos01}. From the results it might
be expected that the quantum state, in which an excited
cavity mode is prepared at some initial time, can be perfectly extracted
from the cavity, so that after sufficiently long time, i.e.,
when the cavity is effectively empty, the pulse which has left the
cavity is in the same quantum
state as the cavity mode initially was. However, in the
analysis the effects of unwanted losses 
(such as absorption and scattering losses)
on the extracted quantum state have not been considered.
Moreover, instead of calculating the quantum state of the outgoing
field directly, the authors base the derivation on an operational
definition of the Wigner function in terms of collective mode 
operators introduced within the frame of the homodyne detection
scheme considered. The reason is that they claim that due to
the mode continuum outside the cavity the Wigner function
would be ill defined. 

In this paper, we directly calculate
as a function of time the quantum state of the pulse which  
leaves a high-$Q$ cavity and may be used for further processing.
The calculations are performed for arbitrary $s$-parametrized
phase space functions, including the Wigner function.
Taking into account both transmission and 
unwanted losses of the cavity mode, we show that
the crucial parameter for the efficiency of
quantum state extraction is the ratio of absorption losses
to transmission losses of the cavity mode. As we will see,
a quantum state can be almost perfectly extracted after sufficiently
long time, only if the value of this ratio
is sufficiently small, thereby the truly required smallness sensitively
depending on the nonclassical features of the state.    

The outline of the paper is as follows. In Sec.~\ref{sec2}
the model is explained and the basic equations, including the
operator input-output relations, are given. The quantum state
of the outgoing field is calculated in Sec.~\ref{sec3}, and
Sec.~\ref{sec4} presents two examples. Finally, a summary
and some concluding remarks are given in Sec.~\ref{sec5}.   


\section{Basic equations}
\label{sec2}

\subsection{Quantum Langevin equation}
\label{sec2.1}

Let us consider a one-dimensional high-$Q$ cavity bounded 
with a perfectly reflecting mirror at $x$ $\!=$ $\!0$ and
an almost perfectly reflecting mirror at $x$ $\!=$ $\!l$.
For a \mbox{high-$Q$} cavity, the
widths $\gamma_k$ of the cavity modes
at frequencies \mbox{$\omega_k$ $\!=$
$\!k\pi c/l$} are very small compared with their separation
\mbox{$\Delta\omega$ $\!=$ $\!\omega_{k+1}$ $\!-$ $\!\omega_k$ $\!=$
$\!\pi c/l$}, where $c$ is the velocity of light.   
Being interested in resolving times that are large compared with
the time of propagation of light
through the cavity,
we may expand the intracavity field in terms of standing
waves at frequencies $\omega_k$, where the associated 
photon creation and annihilation operators $\hat{a}_k ^{\dagger}$ 
and $\hat{a}_k$, respectively, obey quantum Langevin
equations \cite{Gardiner85,Knoell91}.
For sufficiently large $Q$ values, we may further
assume that the time of excitation and preparation of
a cavity wave in a (desired) quantum state is short compared
with its decay time (but still long compared with
the propagation time through the cavity). In this case,
the process of preparation of the cavity quantum state
is well separated from the process
of its transmission to the outside space.      

Let $\hat{\varrho}_\mathrm{cav}$ be the quantum
state an excited cavity wave is prepared in  
at some initial time $t_0$. For times \mbox{$t$ $\!\ge$ $\!t_0$},
the corresponding Langevin equation for the photon
annihilation operator associated with the excited mode then reads     
\begin{eqnarray}
    \label{1}
\lefteqn{
        \dot{\hat{a}}
        = - i
        \left[
        \omega_\mathrm{cav}
        - {\textstyle\frac{i}{2}}
        \left(
        \gamma_\mathrm{rad} + \gamma_\mathrm{abs}
        \right)
        \right]
        \hat{a}
}\nonumber\\[1ex]&&\hspace{2ex}
        + \left(\frac{c}{2l}\right)^{1/2}T \hat{b}_\mathrm{in} (t) 
        + \left(\frac{c}{2l}\right)^{1/2} A \hat{c}(t).  
\end{eqnarray}
In the first term,
\begin{equation}
\label{2}
        \gamma_\mathrm{rad}  = \frac{c}{2l}  |T|^2
\end{equation}
is the decay rate of the cavity mode which
results from the transmission losses due to the radiative
input-output coupling, and
\begin{equation}
\label{3}
        \gamma_\mathrm{abs}  = \frac{c}{2l}  |A|^2
\end{equation}
is the decay rate which 
results from the unwanted losses,
briefly referred to as absorption losses in the rest of
the paper, such as the unavoidably existing material
absorption and scattering.
For a high-$Q$ cavity, both the transmission coefficient $T$
and the absorption coefficient $A$
are very small compared with unity (\mbox{$|T|$ $\!\ll$ $\!1$},
\mbox{$|A|$ $\!\ll$ $\!1$}). Note that $T$ and $A$
are taken at the cavity-mode frequency
$\omega_\mathrm{cav}$. The second term in Eq.~(\ref{1}) is the
Langevin noise force arising from the input radiation field,
where
\begin{eqnarray}
\label{4}
\lefteqn{
        \hat{b}_\mathrm{in} (t) =
        \frac{1} {\sqrt{2\pi}}
        \int_{\Delta\omega}
        \D\omega \, \hat{b}_\mathrm{in} (\omega, t)
}\nonumber\\[1ex]&&
        =\frac{1} {\sqrt{2\pi}}
        \int_{\Delta\omega}
        \D\omega \,\hat{b} (\omega, t_{0})\, e^{-i \omega (t-t_{0})},
\end{eqnarray}
and the third term is the Langevin noise force associated with
absorption, where
\begin{eqnarray}
\label{5}
\lefteqn{
        \hat{c}(t) =
        \frac{1} {\sqrt{2\pi}}
        \int_{\Delta\omega}
        \D\omega \, \hat{c} (\omega, t)
}\nonumber\\[1ex]&&
        =\frac{1} {\sqrt{2\pi}}
        \int_{\Delta\omega}
        \D\omega \, \hat{c} (\omega, t_{0})\, e^{-i \omega (t-t_{0})}.
\end{eqnarray}
Here and in the following, the notation $\int_{\Delta\omega}
\D\omega\ldots$ is used to indicate that the integration runs
over frequencies in the interval \mbox{$[\omega_\mathrm{cav}$
$\!-$ $\!\Delta \omega/2,\,
\omega_\mathrm{cav}$ $\!+$ $\!\Delta \omega/2]$}.
The operators $\hat{a}(t)$, $\hat{b}(\omega,t)$, and $\hat{c}(\omega,t)$
satisfy the familiar bosonic equal-time commutation relations
\begin{equation}
\label{6}
        \left[\hat{a}
        (t),\hat{a}
        ^\dagger(t)\right] =1,
\end{equation} 
\begin{equation}
\label{7}
        \bigl[\hat{b}_\mathrm{in}(\omega,t),
        \hat{b}_\mathrm{in}^\dagger(\omega',t)\bigr] 
        = \delta(\omega-\omega'),
\end{equation}
\begin{equation}
\label{7-1}
        \left[\hat{c}(\omega,t),\hat{c}^\dagger(\omega',t)\right] 
        = \delta(\omega-\omega').
\end{equation}
It is not difficult to see that the solution of Eq.~(\ref{1})
can be given in the form of
\begin{eqnarray}
\label{9}
\lefteqn{
        \hat{a}(t)\,=\,\hat{a}(t_0)
        e^{-i\left[\omega_\mathrm{cav}-\frac{i}{2}({\gamma_\mathrm{rad}
        +\gamma_\mathrm{abs}})\right](t-t_0)}
}\nonumber\\[1ex]&&
        + \left(\frac{c}{2l}\right)^{1/2} \!T \!
        \int_{t_0}^{t}\D t'\,
        e^{-i\left[\omega_\mathrm{cav}-\frac{i}{2}({\gamma_\mathrm{rad}
        +\gamma_\mathrm{abs}})\right](t-t')}\hat{b}_{in}(t')
\nonumber\\[1ex]&&
        + \left(\frac{c}{2l}\right) ^{1/2} \!A\!
        \int_{t_0}^{t}\D t'\, e^{-i\left[\omega_\mathrm{cav}-\frac{i}{2}
        (\gamma_\mathrm{rad}+\gamma_\mathrm{abs})\right](t-t')}\hat{c} (t').
\quad
\end{eqnarray}


\subsection{Input-output relation}
\label{sec2.2}

In close analogy to Eq.~(\ref{4}), output operators
\begin{eqnarray}
\label{10}
\lefteqn{
        \hat{b}_\mathrm{out} (t) =
        \frac{1} {\sqrt{2\pi}}
        \int_{\Delta\omega}
        \D\omega \, \hat{b}_\mathrm{out} (\omega, t)
}\nonumber\\[1ex]&&
        =\frac{1} {\sqrt{2\pi}}
        \int_{\Delta\omega}
        \D\omega \,\hat{b} (\omega, t_{1})\, e^{-i \omega (t-t_1)}
\quad
        (t < t_1)
\quad
\end{eqnarray}
can be introduced, where, similar to Eq.~(\ref{7}), the
bosonic commutation relation
\begin{equation}
\label{11}
\bigl[\hat{b}_\mathrm{out}(\omega,t),
        \hat{b}_\mathrm{out}^\dagger(\omega',t)\bigr]  
        = \delta(\omega-\omega')
\end{equation} 
is valid.
Taking into account that, on the time scale under consideration, 
the lower and upper integration limits of the frequency
integrals can be extended, with little error, to $-\infty$ and $+\infty$,
respectively, from Eqs.~(\ref{4}) and (\ref{10})
together with the commutation relations (\ref{7})
and (\ref{11}) it then follows that the commutation relations
\begin{equation}
\label{8}
        \bigl[\hat{b}_\mathrm{in}(t),\hat{b}_\mathrm{in}^\dagger(t')\bigr] 
        = \delta(t-t')
\end{equation}
and
\begin{equation}
\label{12}
        \bigl[\hat{b}_\mathrm{out}(t),
        \hat{b}_\mathrm{out}^\dagger(t')\bigr] 
        = \delta(t-t')
\end{equation}
may be regarded as being valid. In a similar way, from Eqs.~(\ref{5})
and (\ref{7-1}) we derive
\begin{equation}
\label{8-1}
        \left[\hat{c}(t),\hat{c}^\dagger(t')\right] 
        = \delta(t-t').
\end{equation}
Other important commutation rules are
\begin{equation}
\label{8-2}
\bigl[\hat{a}(t),
        \hat{b}_\mathrm{in}(t')\bigr]
=\bigl[\hat{a}^\dagger(t),
        \hat{b}_\mathrm{in}(t')\bigr]
        =0
\quad \mathrm{if}\quad t<t'.        
\end{equation}

The output operator $\hat{b}_\mathrm{out}(t)$ can be related
to the cavity operator $\hat{a}(t)$ and the input operator
$\hat{b}_\mathrm{in}(t)$ according to the input-output relation
\begin{equation}
\label{13}
        \hat{b} _\mathrm{out} (t) = \left(\frac{c}{2l}\right)^{1/2} T
        \hat{a}
        (t)  + R \hat{b} _\mathrm{in} (t) ,
\end{equation}
where
\begin{equation}
\label{13-1}
R = -\frac {T}{T^*}\,.
\end{equation}
We renounce to repeat its derivation here, but refer the reader to
the literature \cite{Gardiner85,Knoell91}. In Ref.~\cite{Gardiner85}
the derivation of Eq.~(\ref{13}) (with real $T$) is based on 
quantum noise theories whereas in Ref.~\cite{Knoell91} a more rigorous
QED derivation is given (also see Ref.~\cite{Vogel94}).
Equation (\ref{13}) corresponds
to the following equation for the continuous-mode output operators
$\hat{b}_\mathrm{out}(\omega,t)$: 
\begin{eqnarray}
\label{14}
\lefteqn{
        \hat{b}_\mathrm{out}
        (\omega, t) 
}\nonumber\\[1ex]&&
        \, = \,
        \left(\frac{c}{2l}\right)^{1/2} T 
        \lim _{t_1 \rightarrow t _+}\,\frac{1} {\sqrt{2\pi}} 
        \int _{t_0}^{t _1}\D t ' \,
        e^{-i\omega (t - t')}\hat{a} (t') 
\nonumber\\[1ex]&&\hspace{2ex}  
        +\, R \hat{b} (\omega, t_0) \, 
        e^{-i\omega (t - t_0)} .
\qquad         
\end{eqnarray}
The proof of this equation is straightforward. 
Substituting Eq.~(\ref{14}) into Eq.~(\ref{10}),
performing the frequency integral as before
(i.e., extending the integration
limits to $\mp\infty$), and recalling Eq.~(\ref{4}), we
exactly arrive at Eq.~(\ref{13}). Note that
$\hat{b}_\mathrm{out}(\omega,t)$ and
$\hat{b}_\mathrm{out}^\dagger(\omega,t)$, as given by Eq.~(\ref{14}),
fulfill the commutation rule (\ref{11}).

We now substitute Eq.~(\ref{9}) together with Eqs.~(\ref{4})
and (\ref{5}) into Eq.~(\ref{14}) to obtain
\begin{equation}
\label{15}
        \hat{b}_\mathrm{out}
        (\omega, t) =
        F^* (\omega, t)
        \hat{a}(t_0) 
        + \hat{B}(\omega,t),
\end{equation}
where the function $F(\omega,t)$ is defined by
\begin{eqnarray}
\label{17}
\lefteqn{
        F (\omega, t) =  
         \frac{i} {\sqrt{2\pi}} 
        \left(\frac{c}{2l}\right)^{1/2} 
        T^* \,
        e^{ i\omega (t-t_0)}
}\nonumber\\[1ex]&&
\times
        \frac {\exp \left\{- i \left[\omega
        \!-\! \omega_\mathrm{cav}\! -\!
        \frac{i}{2}(\gamma_\mathrm{rad}\! +\!\gamma_\mathrm{abs} )\right]
        (t\!-\!t_0) \right\} \!-\! 1} 
        {\omega - \omega_\mathrm{cav}
        - \frac{i}{2}(\gamma_\mathrm{rad} +\gamma_\mathrm{abs} )}\, , 
\nonumber\\&&
\end{eqnarray}
and the operator $\hat{B}(\omega,t)$ is a linear functional
of the operators $\hat{b}(\omega,t_0)$ and $\hat{c}(\omega,t_0)$
according to
\begin{eqnarray}
\label{16}
\lefteqn{       
        \hat{B}(\omega,t)
        =
        \int_{\Delta\omega} \D \omega '
        G^* (\omega, \omega ', t) \hat{b} (\omega ', t _0)
}\nonumber\\[1ex]&&\hspace{7ex}
        +
        \int_{\Delta\omega}\D \omega '
        H^* (\omega, \omega ', t) \hat{c} (\omega ', t _0) .
\end{eqnarray}
Here, the functions $G(\omega,\omega',t)$ and $H(\omega,\omega',t)$,
respectively, are defined by
\begin{equation}
\label{18}
        G (\omega, \omega ', t) =  
        T ^* \xi (\omega, \omega ', t)
        + 
        R ^* e^{i \omega ' (t-t_0)} \delta (\omega - \omega ')
\end{equation}
and
\begin{equation}
\label{19}
        H (\omega, \omega ', t) =  
        A^* \xi (\omega, \omega ', t) ,
\end{equation}
where $\xi (\omega, \omega ', t)$ reads
\begin{eqnarray}
\label{20}
\lefteqn{
        \xi (\omega, \omega ', t) =  
        \frac {1}{2\pi}\frac{c}{2l}\,T ^{*}
        \frac {1}{
        \omega - \omega_\mathrm{cav} - 
        \frac{i}{2}(\gamma_\mathrm{rad} + \gamma_\mathrm{abs}) }
}\nonumber\\[1ex]&&
\times\,
        \left\{
        \frac {e^{i\omega' (t-t_0)} 
        - e^{i\left[\omega_\mathrm{cav} + \frac
        {i}{2}(\gamma_\mathrm{rad} + \gamma_\mathrm{abs})\right](t-t_0)}} 
                {
        \omega' - \omega_\mathrm{cav} -\frac{i}{2}(\gamma_\mathrm{rad}
        + \gamma_\mathrm{abs}) 
                }
\right.
\nonumber\\[1ex]&&
\left.
\qquad
        -\, 
        \frac {e^{i\omega (t-t_0)} - e^{i\omega' (t-t_0)}} 
                {
        \omega - \omega '
                }
\right\}.
\qquad\qquad
\end{eqnarray}
It is not difficult to see that from Eq.~(\ref{15}) together
with the commutation rules (\ref{6}) and (\ref{11}) it follows that
\begin{equation}
\label{20-1}
        \left[\hat{B}(\omega,t),  \hat{B}^\dagger(\omega',t)  \right] 
        = \delta(\omega-\omega')  - F^*(\omega,t) F(\omega',t).
\end{equation}
Note that
\begin{equation}
\label{20-2}
\bigl[\hat{a}(t_0),\hat{B}(\omega,t)\bigr]
=\bigl[\hat{a}(t_0),\hat{B}^\dagger(\omega,t)\bigr]
= 0.
\end{equation}


\section{Quantum state of the output field}
\label{sec3}


\subsection{Characteristic functional}
\label{sec3.1}

To calculate the quantum state of the output field in the
frequency interval \mbox{$[\omega_\mathrm{cav}$ $\!-$
$\!\Delta \omega/2,\, \omega_\mathrm{cav}$ $\!+$ $\!\Delta \omega/2]$},
we start from its characteristic functional 
\begin{eqnarray}
\label{21}
\lefteqn{
        C_\mathrm{out}[\beta(\omega),t]
}\nonumber\\[1ex]&&
         = \mathrm{Tr}\left\lbrace \hat{\varrho}\, \exp \left[
         \int_{\Delta\omega}
        \!\D\omega\,
         \beta(\omega) \hat{b}^{\dagger}_{out} (\omega, t)
         - \mathrm{H.c.}\right] \right\rbrace,
\quad         
\end{eqnarray}
where $\hat{\varrho}$ is the density operator of the initial quantum
state of the overall system,
i.e., its quantum state at \mbox{$t$ $\!=$ $\!t_0$}.
To further handle the functional,
it is convenient to regard the
integral as the limit of a sum, perform the
calculations for the sum, and take the limit at the
end of the calculations. That is to say, we write   
\begin{equation}
\label{22}
        C_\mathrm{out}[\beta(\omega),t] = \lim_{N\to\infty}
        C_\mathrm{out}(\boldsymbol{\beta},t)
\end{equation}
[$\boldsymbol{\beta}$ $\!\equiv$ $\!(\beta_1,\beta_2,\ldots,\beta_N)$],
where
\begin{equation}
\label{23}
        C_\mathrm{out}(\boldsymbol{\beta},t)
        = \mathrm{Tr} \left\lbrace \hat{\varrho}\,
        \exp\!\left[ 
         \sum _{n=1}^N \beta _n \hat{b} ^{\dagger}_{n} ( t)
         - \mathrm{H.c.}
 \right] \right\rbrace .
\end{equation}
Here, $\beta _n$ and $\hat{b}_{n} ( t)$, respectively, are
defined by
\begin{equation}
\label{24-1}
    \beta _n = \frac{1}{\sqrt{\delta\omega}}
    \int\limits_{\omega_n-\delta\omega/2}^{\omega_n+\delta\omega/2}
    \D\omega\,\beta(\omega)
    = \sqrt{\delta\omega}\,\beta(\omega_n)
\end{equation}
and
\begin{equation}
\label{24}
        \hat {b} _n (t) = \frac{1}{\sqrt{\delta\omega}}
        \int\limits_{\omega_n-\delta\omega/2}^{\omega_n+\delta\omega/2}
        \D\omega\,\hat{b}_\mathrm{out}(\omega,t)
\end{equation}
($\delta\omega$ $\!=$ $\!\Delta\omega/N$). Note that
\begin{equation}
\label{26}
        \bigl[\hat{b}_n(t),\hat{b}_{n'}^\dagger(t)\bigr] = \delta_{nn'}\,.
\end{equation}
The discrete version of Eq.~(\ref{15}) then reads
\begin{equation}
\label{25}
        \hat{b}_{n} (t) =
        F^* _n(t)
        \hat{a}(t_0) + \hat{B}_n (t)\,,
\end{equation}
where $F_n(t)$ and $\hat {B} _n (t)$ being defined according
to Eqs.~(\ref{24-1}) and (\ref{24}), respectively, with
$F(\omega,t)$ and $\hat {B}(\omega,t)$ instead of 
$\beta(\omega)$ and $\hat {b}_\mathrm{out}(\omega,t)$, respectively,
and Eq.~(\ref{20-1}) changes to
\begin{equation}
\label{27}
        \bigl[\hat{B}_n (t),  \hat{B}_{n'} ^{\dagger} (t)  \bigr] 
        = \delta _{n n'}  - F_{n} ^*(t) F_{n'}(t).
\end{equation}

Let us assume that the (initial) density operator $\hat{\varrho}$
is factorable as
\begin{equation}
\label{27-1}
\hat{\varrho} =
\hat{\varrho}_\mathrm{cav}\otimes\hat{\varrho}_\mathrm{in}
\otimes\hat{\varrho}_\mathrm{abs}
\end{equation}
($\hat{\varrho}_\mathrm{cav}$, density operator of the cavity mode;
$\hat{\varrho}_\mathrm{in}$, density operator of the input field;
$\hat{\varrho}_\mathrm{abs}$, density operator of the dissipative
system responsible for absorption).
Substituting Eq.~(\ref{25}) into Eq.~(\ref{23}), we may write,
on recalling Eq.~(\ref{20-2}),
\begin{eqnarray}
\label{27-2}
\lefteqn{
C_\mathrm{out}(\boldsymbol{\beta},t) 
= \mathrm{Tr}\left\lbrace \hat{\varrho}_\mathrm{cav} 
        \exp\!\left[ 
        \sum _{n=1}^N \beta _n
        F _n(t)
        \hat{a} ^{\dagger}(t_0)  
        - \mathrm{H.c.} \right] \right\rbrace 
}
\nonumber\\[1ex]&&\hspace{2ex}\times\,
        \mathrm{Tr} \left\lbrace \hat{\varrho}_\mathrm{in}
        \otimes\hat{\varrho}_\mathrm{abs}
        \exp\!\left[ 
         \sum _{n=1}^N \beta _n \hat{B} ^{\dagger}_n ( t)
         - \mathrm{H.c.}
 \right] \right\rbrace .
\qquad
\end{eqnarray}
In what follows we consider the case in which 
both the input field and the dissipative system are initially
in the vacuum state. In this case, the second trace in
Eq.~(\ref{27-2}) simply reduces
\begin{eqnarray}
\label{28}
\lefteqn{
\mathrm{Tr} \left\lbrace \hat{\varrho}_\mathrm{in}
        \otimes\hat{\varrho}_\mathrm{abs}
        \exp\!\left[ 
         \sum _{n=1}^N \beta _n \hat{B} ^{\dagger}_n ( t)
         - \mathrm{H.c.}
 \right] \right\rbrace
}
\nonumber\\[1ex]&&
= \exp\!\left\lbrace - {\textstyle\frac {1}{2}}  
        \left[\sum _{n=1}^N |\beta _n| ^2 \!-\!\!\! 
        \sum_{n,n'=1}^N \beta _n
        F_n(t)
        \beta ^* _{n'}
        F^* _{n'}(t)
        \right]
 \right\rbrace,
\nonumber\\&&
\end{eqnarray}
which can be easily proved to be correct by recalling the 
commutation rule (\ref{27}) and applying the
Baker-Campbell-Hausdorff formula to write the exponential
operator in normal order.    
Combining Eqs.~(\ref{27-2}) and (\ref{28}), we may rewrite
$C_\mathrm{out}(\boldsymbol{\beta},t)$ as
\begin{eqnarray}
\label{29}
\lefteqn{
        C_\mathrm{out}(\boldsymbol{\beta},t)
}
\nonumber\\[1ex]&&
        = \exp\!\left[- {\textstyle\frac {1} {2}} 
        \sum _{n=1}^N |\beta _n| ^2 \right] 
        e^{{\textstyle\frac {1} {2}} | \lambda(\boldsymbol{\beta},t) | ^2}
        C_\mathrm{cav}[\lambda(\boldsymbol{\beta},t)].
\quad
\end{eqnarray}
Here,
\begin{equation}
\label{30}
        C_\mathrm{cav}(\beta)
        = \mathrm{Tr} \left[ \hat{\varrho} _\mathrm{cav}
        e^{\beta\hat{a}^\dagger(t_0)-\beta^\ast\hat{a}(t_0)}
        \right]
\end{equation}
is the characteristic function of the quantum state of
the cavity mode,
and the function $\lambda(\boldsymbol{\beta},t)$ is defined by
\begin{equation}
\label{31}      
        \lambda(\boldsymbol{\beta},t) = 
        \sum _{n=1}^N  F_n(t) \beta _n .
\end{equation}

Equation (\ref{29}) relates the multidimensional characteristic
function of the quantum state of the
multimode output field (at time \mbox{$t$ $\!\ge$ $\!t_0$}) to the
characteristic function of the cavity-mode quantum state (at time $t_0$).
Let $C_\mathrm{out}(\boldsymbol{\beta},t;s)$ and $C_\mathrm{cav}(\beta;s_0)$
be the respective characteristic functions in
arbitrary $s$- and $s_0$-order, respectively. The
extension of Eq.~(\ref{29}) valid for \mbox{$s$ $\!=$ $s_0$ $\!=$ $\!0$} 
to arbitrary values of $s$ and $s_0$ is straightforward:
\begin{eqnarray}
\label{32}
\lefteqn{
        C_\mathrm{out}(\boldsymbol{\beta},t;s)
        = \exp\!\left[{\textstyle\frac {1} {2}} 
        \sum _{n=1}^N |\beta _n| ^2 (s-1)\right]
}\nonumber\\[1ex]&&\hspace{10ex}
\times\,
        e^{{\textstyle\frac {1} {2}}
        | \lambda(\boldsymbol{\beta},t) | ^2(1-s_0)}
        C_\mathrm{cav}[\lambda(\boldsymbol{\beta},t);s_0].
\qquad          
\end{eqnarray}


\subsection{Phase-space functions}
\label{sec3.2}

Let us now turn from the relation between the characteristic functions
$C_\mathrm{out}(\boldsymbol{\beta},t;s)$ and $C_\mathrm{cav}(\beta;s_0)$
to the relation between the corresponding phase-space functions
\begin{eqnarray}
\label{32-1}
\lefteqn{
P_\mathrm{out}(\boldsymbol{\alpha},t;s)
= \frac{1}{\pi^{2N}} \int \D ^{2N} \beta\,
C_\mathrm{out}(\boldsymbol{\beta},t;s)
}
\nonumber\\[1ex]&&\hspace{14ex}\times\,
\exp\!\left[\sum _{n=1}^N (\alpha _n \beta _n^\ast 
        - \alpha _n^\ast \beta _n)\right],
\qquad        
\end{eqnarray}
and
\begin{equation}
\label{32-2}
P_\mathrm{cav}(\alpha;s_0)
= \frac{1}{\pi^2} \int \D^2\beta\,
C_\mathrm{cav}(\beta;s_0) e^{\alpha\beta^\ast - \alpha^\ast\beta}, 
\end{equation}
respectively.
Taking $C_\mathrm{out}(\boldsymbol{\beta},t;s)$ from Eq.~(\ref{32}),
we derive         
\begin{eqnarray}
\label{34}
\lefteqn{
        P_\mathrm{out}(\boldsymbol{\alpha}, t; s)
        = \frac {1} {\pi ^{2N}} \int \D ^{2N} \beta
        \int \D ^2 \alpha \, P_\mathrm{cav} (\alpha ; s_0)
}
\nonumber\\[1ex]&&
\times\,
        \exp\!\left[  
        \sum _{n=1}^N (\alpha _n \beta _n^\ast 
        - \alpha _n^\ast \beta _n)\right]
        \exp\!\left[ {\textstyle\frac {1} {2}} 
        \sum _{n=1}^N |\beta _n| ^2 (s-1)\right]
\nonumber\\[1ex]&&
\times\,  
        \exp\!\left[ \lambda(\boldsymbol{\beta},t)\alpha ^*
        \!-\! \lambda ^*(\boldsymbol{\beta},t) \alpha  \right]
        \exp\!\left[{\textstyle\frac {1} {2}} 
        | \lambda(\boldsymbol{\beta},t)| ^2 (1\!-\! s_0)\right]\!.
\quad
\nonumber\\[1ex]&&
\end{eqnarray}
To perform the $2N$-fold integral over the $\beta_n$, we change
the variables by means of a unitary transformation
\begin{equation}
\label{35}
        \beta'_m = \sum_{n=1}^N U_{mn} \beta_n\,,
\end{equation}
\begin{equation}
\label{35-1}
\bigl(U^{-1}\bigr)_{mn} = U^\ast_{nm}\,,
\end{equation}
thus
\begin{equation}
\label{35-2}
\beta_n = \sum_{m=1}^N U^\ast_{mn} \beta_m'\,.
\end{equation}
In fact, this transformation
corresponds to the introduction of non-monochromatic modes, the
phase-space variables of which are given by
\begin{equation}
\label{36}
        \alpha'_m = \sum_{n=1}^N U_{mn} \alpha_n\,.
\end{equation} 

In order to diagonalize the quadratic form in the last exponental
in Eq.~(\ref{34}), we set
\begin{equation}
\label{36-1}
U_{1n} = \frac{F_n(t)}{\sqrt{\eta(t)}}\,,
\end{equation} 
where
\begin{equation}
\label{38}
        \eta(t)  = \sum _{n=1}^N |F_n(t)| ^2,
\end{equation}
so that, according to Eq.~(\ref{35}), $\beta_1'$ is expressed
in terms of the $\beta_n$ as
\begin{equation}
\label{37}
        \beta _1 ' = \,  \frac {1} {\sqrt{
        \eta(t)
        }}
        \sum_{n=1}^N
        F_n(t)
        \beta _n\, . 
\end{equation}
In this case, the multimode phase-space function in the new variables,
simply reduces to the product of single-mode phase-space functions
[$P_\mathrm{out}(\boldsymbol{\alpha},t;s)$ $\!\to$
$\!P_\mathrm{out}(\boldsymbol{\alpha}',t;s)$], 
\begin{eqnarray}
\label{39}
\lefteqn{
        P_\mathrm{out}(\boldsymbol{\alpha}',t;s)\!
        =\! P_\mathrm{out}(\alpha'_1,t;s)
        P_\mathrm{out}(\alpha'_2,t;s)\cdots
}
\nonumber\\[1ex]&&\hspace{10ex}
        \cdots P_\mathrm{out}(\alpha'_{N-1},t;s)
        P_\mathrm{out}(\alpha'_N,t;s),
\quad
\end{eqnarray}
as it is easily seen from Eq.~(\ref{34}).
Obviously, only the first of these output modes is related to
the cavity mode, whereas all other modes are in the
vacuum state. From Eq.~(\ref{34}) it then follows that
the phase-space function of the relevant
output mode is given by (\mbox{$\alpha '$ $\!\equiv$ $\!\alpha_1'$},
\mbox{$\beta'$ $\!\equiv$ $\!\beta_1'$})
\begin{eqnarray}
\label{40}
\lefteqn{ 
        P_\mathrm{out}(\alpha',t;s)
        = \frac {1} {\pi ^2}
        \int \D ^2 \beta' \int \D ^2 \alpha \,
        P _\mathrm{cav} (\alpha;s_0)
}
\nonumber\\[1ex]&& \times\,
        \exp \Bigl\{
        \left[\sqrt{
        \eta(t)
        }\, \alpha ^* 
        - \alpha'{^*} \right]\beta'
        -\left[
        \sqrt{
        \eta(t)
        }\, \alpha      
        - \alpha' \right]\beta'{^\ast}
\Bigr. \nonumber\\[1ex]&&\hspace{10ex}
        \Bigl.
        -\,{\textstyle\frac {1}{2}}\left[1-s-
        \eta(t)\,
        (1-s_0)
        \right]
        |\beta'|^2  
        \Bigr\},
\end{eqnarray}
which after integration over $\beta'$ yields
\begin{eqnarray}
\label{41}
\lefteqn{
        P_\mathrm{out}( \alpha',t ; s )
        = \frac {2} {\pi}
        \frac {
        1
        } {1- s -
        \eta(t)
        (1-s_0)} 
}\nonumber\\[1ex]&&
\times  
        \int \D ^2 \alpha\,  P _\mathrm{cav} (\alpha; s_0)  
        \exp\!\left[ -\frac {2|\sqrt{
        \eta(t)
        }\, \alpha \! -\!\alpha'|^2 }
        {1\!-\! s \!-\!
        \eta(t)
        (1\!-\! s_0)}   \right]\!,\qquad
\end{eqnarray}
provided that
\begin{equation}
\label{42}
        1- s -
        \eta(t)
        (1- s_0) \ge 0.
\end{equation}
Note that the case of equality sign 
should be understood as limiting process. 
We compare Eq.~(\ref{41}) with the well-known relation
\begin{equation}
\label{43}
        P(\alpha ; s )
        = \frac {2} {\pi (s'-s)}
        \int \D ^2 \beta \,  P(\beta; s ' )  
        \exp\!\left[ -\frac {2|\beta -\alpha|^2 }{s'-s}   \right],
\end{equation}
which is valid for
\begin{equation}
\label{43-1}
        s'-s\ge 0 ,
\end{equation}
and see that the quantum state of the relevant output mode
can be expressed in terms of the quantum state of the cavity mode
in the compact form of 
\begin{equation}
\label{44}
        P_\mathrm{out}(\alpha,t;s)
        =\, \frac {1} {\eta (t)}\, 
        P _\mathrm{cav} \!\left[\frac{\alpha}{\sqrt{
        \eta(t)
        }} ; s ' \right], 
\end{equation}
where, for chosen value of $s$, the value of $s '$ is given by   
\begin{equation}
\label{45}
        s ' = 1- \frac{1-s} {\eta(t)} .
\end{equation}
To calculate $\eta(t)$, we recall that
in the limit $N$ $\!\to$ $\!\infty$ 
\begin{equation}
\label{46}
        \eta(t)  = \lim_{N\to\infty}\sum_{n=1}^N | F_n(t)| ^2 =
        \int_{\Delta\omega}
        \D\omega \,
        |F(\omega,t)| ^2 ,
\end{equation}
with $F(\omega,t)$ from Eq.~(\ref{17}). Straightforward
calculation yields
\begin{equation}
\label{47}
        \eta(t) = \frac {\gamma_\mathrm{rad} }
        {\gamma_\mathrm{rad} + \gamma_\mathrm{abs}}
        \left[ 1- e^{- (\gamma_\mathrm{rad} +
        \gamma_\mathrm{abs}) (t-t_0)}\right].
\end{equation}
\begin{figure}[tb]
 \includegraphics
 {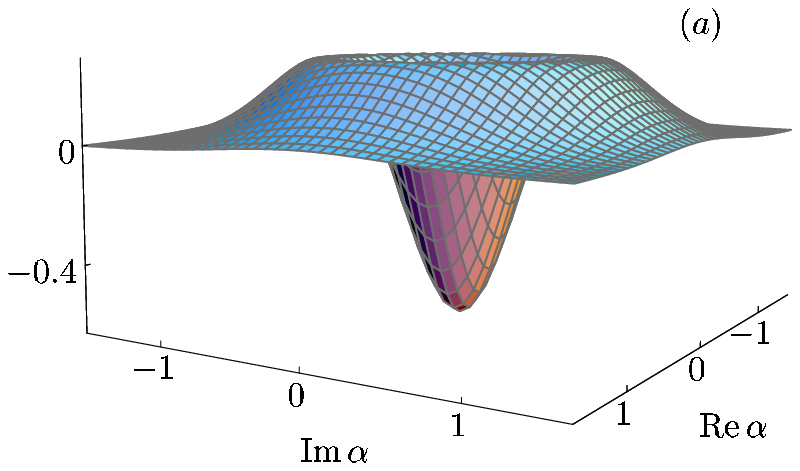}
 \vspace{8mm}\\
 \includegraphics
 {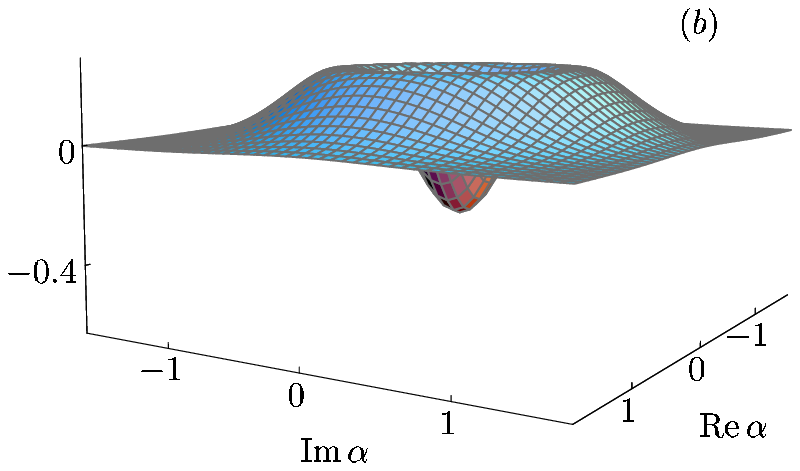}
 \vspace{8mm}\\
 \includegraphics
 {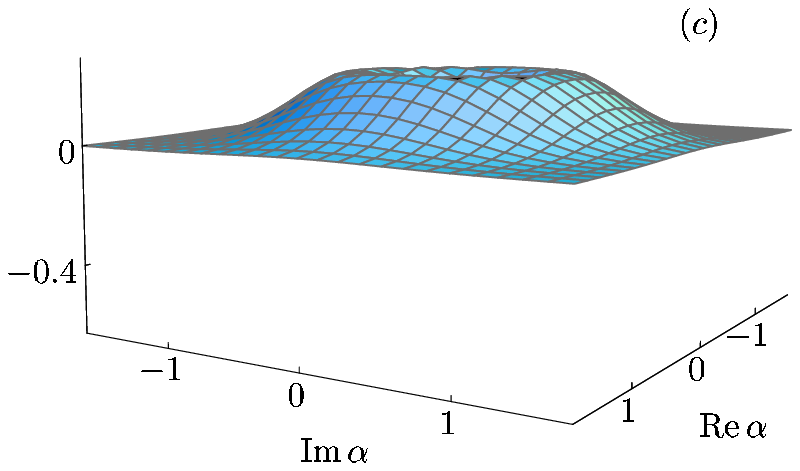}
\caption{\label{fig1}
Wigner function of the quantum state of the pulse that
leaves the cavity,
the mode of which is (initially) prepared
in a single-photon state.
(a) $\eta(t)$ $\!=$ $\!0.99$;
(b) $\eta(t)$ $\!=$ $\!0.71$;
(c) $\eta(t)$ $\!=$ $\!0.5$.
}
\end{figure}

Setting in Eq.~(\ref{41}) \mbox{$s$ $\!=$ $\!s_0$ $\!=$ $\!0$}, we
see that the Wigner function of the relevant output mode is the following
convolution of the Wigner function of the cavity mode with a Gaussian:  
\begin{eqnarray}
\label{48}
\lefteqn{
        W_\mathrm{out}(\alpha,t)
        = \frac {2} {\pi}\frac {
        1
        } {1-
        \eta(t)
        }
}\nonumber\\*[1ex]&&
\;\times \, 
         \int \D ^2 \beta\,
        W _\mathrm{cav} (\beta) \, 
        \exp\!\left[ -\frac {2|\sqrt{
        \eta(t)
        } \beta  -\alpha|^2 }
        {1-
        \eta(t)
        } \right]\! .
\quad        
\end{eqnarray}
This equation reveals that
for perfectly extracting a quantum state from a
high-$Q$ cavity, the condition
\begin{equation}
\label{49}
        \frac{
        \eta(t)
        }{1-
        \eta(t)
        } \gg 1
\end{equation}
should be satisfied, i.e., the value of the extraction
efficiency $\eta(t)$ must be sufficiently close
to unity. How close to unity -- it really depends on
the characteristic quantum features of the state to be extracted.
On the other hand, from Eq.~(\ref{47})
it follows that
\begin{equation}
\label{50}
        \eta(t)
        \leq \frac{\gamma_\mathrm{rad}}
        {\gamma_\mathrm{rad}+\gamma_\mathrm{abs}}\,.
\end{equation}
Note that \mbox{$
\eta(t)
$ $\!\simeq$ $\!\gamma_\mathrm{rad}/
(\gamma_\mathrm{rad}$ $\!+$ $\!\gamma_\mathrm{abs})$}
for sufficiently long times
\mbox{$t$ $\!-$ $\!t_0$ $\!\gtrapprox$ $\!(\gamma_\mathrm{rad}$
$\!+$ $\!\gamma_\mathrm{abs})^{-1}$}.
 

\section{Examples}
\label{sec4}

The really required efficiency for nearly perfect
quantum state extraction sensitively depends on the
quantum state that is desired to be extracted. To
illustrate this, let us consider two examples of
highly nonclassical states, namely Fock states and
Schr\"{o}dinger catlike states.


\subsection{Fock states}

A typical nonclassical state is an $n$-photon
Fock state, whose Wigner function reads 
\begin{equation}
\label{51}
        W_\mathrm{cav} ^{(n)} ( \alpha )
        =   \frac {2 } {\pi}
        (-1)^n
        e^{-2 |\alpha |^2}
        \mathrm{L}_n\! \left(4 |\alpha |^2\right) , 
\end{equation}
where \mbox{$\mathrm{L}_n (x)$} is the Laguerre polynomial
of order $n$. 
Substituting Eq.~(\ref{51}) into Eq.~(\ref{48}) and
employing the integral representation of 
the Laguerre polynomials \cite{Arfken},
\begin{eqnarray}
\label{52-1}
        \mathrm{L}_n\! \left(x\right) = 
        \frac {1} {2\pi i} 
        \oint _{\gamma}
        \D z \,
        \frac{
        e^{-x z/(1-z)}
        }{
        z^{n+1}(1-z)} \,,
\end{eqnarray}
where the contour $\gamma$ encloses the origin but not the point 
$z = 1$, after straightforward calculations we obtain the 
Wigner function of the output pulse as
\begin{eqnarray}
\label{52-0}
\lefteqn{
        W_\mathrm{out}^{(n)}(\alpha,t) = 
                \frac {2 } {\pi} 
                (-1)^n \,
                e^{-2 |\alpha |^2}      
                \left[ 2 \eta (t) - 1 \right] ^n
}\nonumber\\*[1ex]&&\hspace{10ex}
\times\,  
        \mathrm{L}_n\! \left[
                \frac{4 \eta (t)}{2\eta (t) - 1 }
                |\alpha |^2\right] .
\end{eqnarray}
From Eq.~(\ref{52-0}) it is not difficult to see that the condition  
\begin{equation}
\label{53}
        \eta(t) > 1- \frac {1} {2 n}
\end{equation}
must be satisfied to guarantee that the 
$n$-photon Fock state prevails in the
mixed output quantum state.
In the simplest case 
of a one-photon Fock state,
\mbox{$n$ $\!=$ $\!1$}, 
the condition reduces to $\eta(t)$ $\!>$ $\!0.5$.
That is to say, the weight of the one-photon Fock state
exceeds the weight of the vacuum state 
in the mixed state of the outgoing field,
\begin{equation}
\label{69-1}
W_\mathrm{out}^{(1)}(\alpha,t)
= [1-\eta(t)]W^{(0)}(\alpha) + \eta(t)W^{(1)}(\alpha),
\end{equation}
only if the extraction efficiency exceeds $50\%$.
The condition (\ref{53}) clearly shows that with increasing value of $n$
the required extraction efficiency rapidly approaches $100\%$.  

The dependence on the extraction efficiency of the quantum state
of the outgoing field is illustrated in Fig.~\ref{fig1}
for the case in which a single-photon Fock state is desired to
be extracted. Figure \ref{fig1}(a) reveals that nearly perfect
extraction requires an extraction efficiency that should
be not smaller than $\eta(t)$ $\!=$ $\!0.99$, which
for $t$ $\!\to$ $\!\infty$ corresponds to the requirement that
$\gamma_\mathrm{abs}/\gamma_\mathrm{rad}$ $\!\lesssim$ $\!0.01$.
As long as \mbox{$\eta(t)$ $\!>$ $\!0.5$}, the single-photon Fock
state is the dominant state in the mixed output state, as
can be seen from Fig.~\ref{fig1}(b) [$\eta(t)$ $\!=$ $\!0.71$,
i.e., $\gamma_\mathrm{abs}/\gamma_\mathrm{rad}$ $\!=$ $\!0.429$
\mbox{($t$ $\!\to$ $\!\infty$)}].
For $\eta(t)$ $\!\leq$ $\!0.5$,
i.e., $\gamma_\mathrm{abs}/\gamma_\mathrm{rad}$ $\!\ge$ $\!1$
($t$ $\!\to$ $\!\infty$)], the features typical of a
single-photon Fock state are 
lost, Fig.~\ref{fig1}(c).


\subsection{Schr\"odinger catlike states}

\begin{figure}[!t!]
 \includegraphics
 {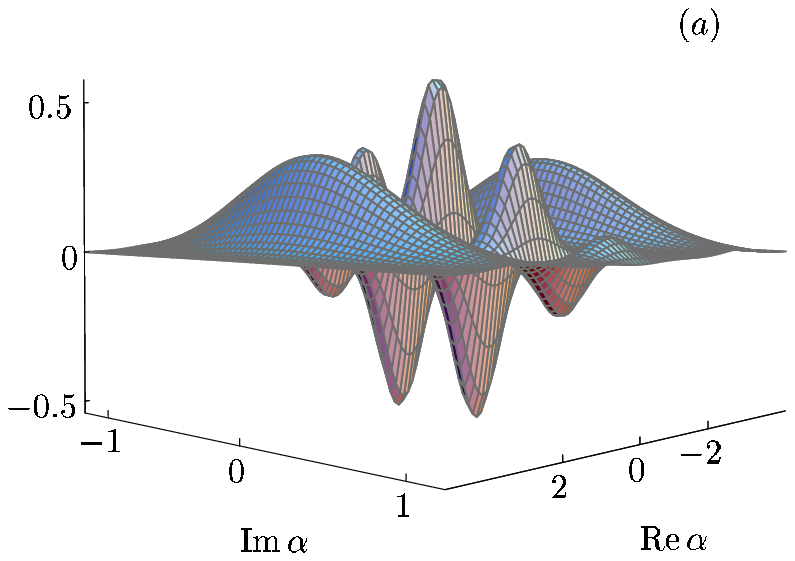}
 \vspace{8mm}\\
 \includegraphics
 {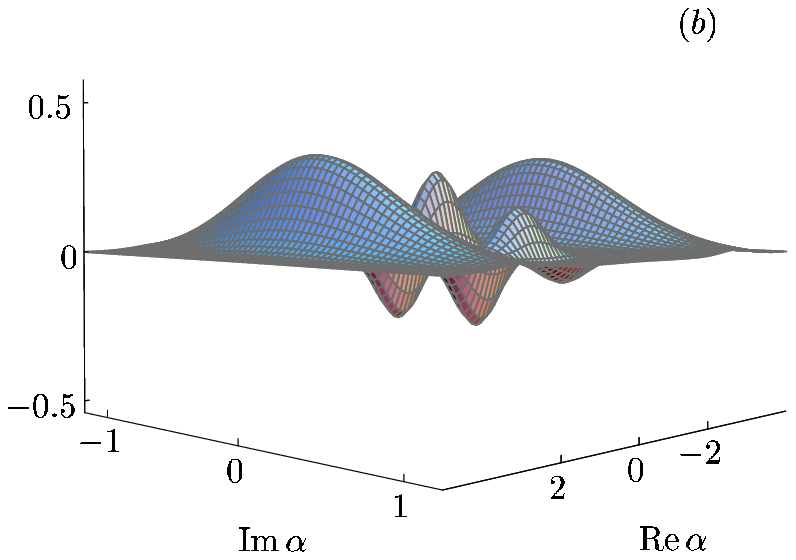}
 \vspace{8mm}\\
 \includegraphics
 {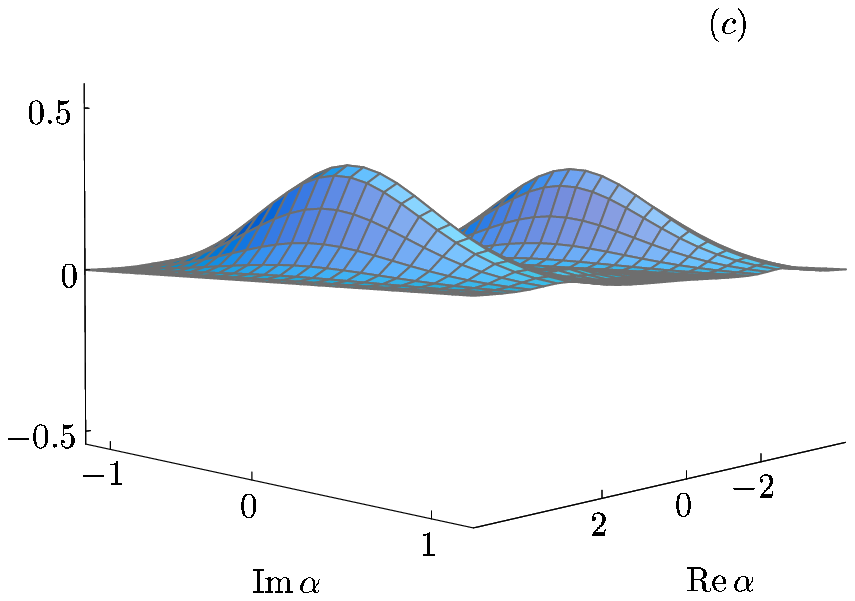}
\caption{\label{fig2}
Wigner function of the quantum state of the pulse that
leaves the cavity,
the mode of which is (initially) prepared
in a Schr\"o\-dinger catlike state given by
Eq.~(\ref{54}) with $\alpha_0$ $\!=$ $\! 3$.
(a) $\eta(t)$ $\!=$ $\!0.998$;
(b) $\eta(t)$ $\!=$ $\!0.952$;
(c) $\eta(t)$ $\!=$ $\!0.84$.
}
\end{figure}

Another example of typically nonclassical states are
Schr\"odinger catlike states, e.g.,
\begin{equation}
\label{54}
        \left| \psi \right\ket_\mathrm{cav}
        =  \mathcal{N} \left( \left| \alpha_0 \right\ket  +
        \left| -\alpha_0 \right\ket\right) ,
\end{equation}
with $\alpha_0$ real, and
\begin{equation}
\label{55}
        \mathcal{N}
        = \left[2\left(1+\, e^{-4\alpha_0 ^2 }\right)\right] ^{-1/2}\,.
\end{equation}
The Wigner function of such a state is given by
\begin{eqnarray}
\label{56}
\lefteqn{
        W_\mathrm{cav} ( \alpha )
        =  \frac {2\mathcal{N}^2} {\pi}
        \left[ e^{-2 |\alpha -\alpha_0 |^2}
\right.
}\nonumber\\*[1ex]&&
\left.
        +\, e^{-2 |\alpha +\alpha_0 |^2}
        + 2 \, e^{-2 |\alpha|^2} \cos\!\left(4 \alpha_0
        \mathrm{Im}\,\alpha  \right) \right] . 
\end{eqnarray}
Substitution of Eq.~(\ref{56}) into Eq.~(\ref{48}) yields the
following expression for the Wigner function of the output fields:
\begin{eqnarray}
\label{57}
\lefteqn{
        W_\mathrm{out} ( \alpha,t )
        = \frac {2\mathcal{N}^2
        } {\pi}
}\nonumber\\*[.5ex]&&
\times\,
        \left\{
        e^{-2 |\alpha - \sqrt{
        \eta(t)
        }\alpha_0 |^2}
        + e^{-2 |\alpha  + \sqrt{
        \eta(t)
        }\alpha_0 |^2}
\right.
\nonumber\\[.5ex]&&
\left.
        +\, 2 e^{-2|\alpha|^2}
        \cos\left[4 \sqrt{
        \eta(t)
        }\alpha_0
        \mathrm{Im}\,\alpha  \right]
        e^{-2 \alpha _0 ^2 [1-
        \eta(t)
                ]}
        \right\} .
\quad
\end{eqnarray}
From Eq.~(\ref{57}) it follows that nearly perfect extraction
of the state requires the condition
\begin{equation}
\label{58}
        1-\eta(t) \ll \frac{1}{2|\alpha_0|^2}
\end{equation}
to be satisfied.

Figure \ref{fig2} illustrates the 
dependence on the extraction efficiency of the quantum state
of the outgoing field for a Schr\"o\-dinger catlike cavity
state with \mbox{$\alpha_0$ $\!=$ $\!3$}.
Comparing Fig.~\ref{fig2} with Fig.~\ref{fig1}, we see
that, as expected, the efficiency
for extracting such a Schr\"{o}dinger catlike state
is required to be substantially higher
than that for extracting a single-photon Fock state. 
For a nearly perfect extraction of
the chosen Schr\"{o}dinger catlike state, the efficiency should be
not smaller than $\eta(t)$ $\!=$ $\!0.998$, i.e.,
$\gamma_\mathrm{abs}/\gamma_\mathrm{rad}$ $\!\lesssim$ $\!0.002$
for $t$ $\!\to$ $\!\infty$ [Fig~\ref{fig2}(a)].
The nonclassical interference fringes typical of a Schr\"{o}dinger
catlike state can be observed, at least rudimentarily, as long as
\mbox{$\eta(t)$ $\!>$ $\!0.84$}, i.e.,
\mbox{$\gamma_\mathrm{abs}/\gamma_\mathrm{rad}$ $\!<$ $\!0.19$}
\mbox{($t$ $\!\to$ $\!\infty$)} [Fig~\ref{fig2}(b); 
\mbox{$\eta(t)$ $\!=$ $\!0.952$, i.e.,$\gamma_\mathrm{abs}
/\gamma_\mathrm{rad}$ $\!=$ $\!0.05$}
\mbox{($t$ $\!\to$ $\!\infty$)}]. 
For smaller values of the extraction efficiency, the quantum
interferences are effectively destroyed [Fig.~\ref{fig2}(c)].


\section{Summary and Conclusions}
\label{sec5}

We have derived an input-output relation that
relates the quantum state of the pulse leaving a high-$Q$
cavity to the quantum state in which an excited cavity mode
was prepared at some initial time. Performing the
calculations in the phase space, we have represented the
respective quantum states in terms of $s$-parametrized phase-space
functions and derived a formula that relates the
phase-space functions of the outgoing field and the
cavity mode to each other. Taking into account unwanted
losses of the cavity mode, we have studied the conditions
under which a nearly perfect extraction of non-classical
quantum states from high-$Q$ cavities should be possible. 

To calculate the quantum state of the outgoing field,
we started from its time-dependent continuous multimode
characteristic functional. By appropriate diagonalization,
it can be rewritten in terms of non-mono\-chro\-matic modes, 
one of which is related to the cavity mode, while all other
modes remain unaffected by the cavity mode. In this
way, the $s$-parametrized phase-space functions of the
quantum state of the relevant non-monochromatic output mode
can be expressed in terms of $s$-parametrized phase-space
functions of the quantum state in which the cavity mode
was prepared. In particular, the output Wigner function can be
given as a convolution of the cavity Wigner function
with a Gaussian reflecting the unwanted losses.       

The crucial parameter for nearly perfect extraction of
a quantum state from a high-$Q$ cavity is the
extraction efficiency,
which in the long-time limit
is determined by the ratio between the cavity-mode decay rate
due to unwanted losses and the cavity-mode decay rate due
to wanted (i.e., transmission) losses. 
This ratio must be sufficiently small in order to
realize a nearly $100\%$ extraction efficiency, where the really required
smallness sensitively depends on the quantum state to be exctracted.  In
particular, extracting highly nonclassical states can require extremely small
values of this ratio.

It should be pointed out that even for the best optical
high-$Q$ microcavities available the required efficiencies for
nearly perfect extraction of nonclassical quantum states
have not been reached, because the unwanted losses are of
the same order of magnitude as the transmission losses
\cite{Rempe92,Hood01,Pelton02}.
So, in the simplest case of extracting from a cavity
a one-photon Fock state, the weight of the one-photon
Fock state exceeds the weight of the vacuum state in
the mixed output quantum state only if
the extraction efficiency is bigger than $50\%$.
However, the biggest value that has been realized
so far in the production of triggered single photons by coupling
a single semiconductor quantum dot to an optical mode in a 
micropost microcavity is about 
$38\%$ \cite{Pelton02}. On the contrary, in
case of high-$Q$ microwave cavities the absorption losses may be small
compared with the transmission losses \cite{Walther03}.

We have concentrated on the
calculation of the quantum state of the field
that leaves a single cavity that is initially excited
in some single-mode quantum state. The theory can also
be extended to multimode excitation in a single
cavity as well as multi-cavity systems.
We have further assumed that the input field is in the vacuum
quantum state. Clearly, the underlying formalism can also
be applied to the case, in which the input field is prepared
in another than the vacuum state. Needless to say that  
when the input field is in a thermal state, then additional
noise is fed into the cavity, and the quantum state of the
output field also carries additional noise.    
As can be seen from Eq.~(\ref{13}), the operator input-output
relation used in this paper
does not take into account that the input field could be
absorbed in the entrance port of the cavity, which would also
give rise to additional noise. To include this effect in the
theory, the input-output relation (\ref{13}) should be
generalized, e.g., by following the line
in Ref.~\cite{Khanbekyan03}.

Finally, we have assumed that the process of preparation of
the quantum state of the cavity mode is sufficiently short
compared with the decay time of the cavity mode, so that
the time scales of quantum state preparation and extraction
are well separated from each other and the preparation process
can be ignored in the calculations. At this point it
should be mentioned that one possible way
to reduce the effect of unwanted losses may be the use
of cavities of deliberately enlarged transmission, so that
the unwanted losses become small compared with transmission
losses. When, for example, the radius of a microsphere
cavity is diminished, then the transmission losses increase,
thereby the absorption losses remaining nearly constant   
\cite{Ho01,Buck03}. Since, on the other hand, the quality factor
is reduced, the preparation time may be comparable
with the cavity decay time, which is now determined by
the transmission time. So, in the
single-photon emitter experiments in Ref.~\cite{Kuhn02},
in which a cavity of a $Q$ value of $6\!\times\!10^4$ is used, the
measured transmission time of several microseconds
is of the same order of magnitude as the cavity decay time.
In order to answer the question of which
quantum state is really obtained
outside the cavity in such a case, the preparation process
must necessarily be included in the calculations.   


\acknowledgments 
M.K. and D.-G.W. would like to thank 
Christian  Raabe and Stefan Scheel for valuable discussions.
A.A.S. and W.V. gratefully acknowledge
support by the Deutsche Forschungsgemeinschaft.



\begin{references}

\bibitem{Raimond01}
J. M. Raimond, M. Brune, and S. Haroche,
Rev. Mod. Phys. {\bf 73}, 565 (2001).

\bibitem{Pinkse00}
P. W. H. Pinkse, T. Fischer, P. Maunz, and G. Rempe,
Nature {\bf 404}, 365 (2000).

\bibitem{Hood00}
C. J. Hood, T. W. Lynn,  A. C. Doherty,  A. S. Parkins,
and H. J. Kimble,
Science {\bf 287}, 1447 (2000).

\bibitem{Doherty00}
A. C. Doherty, T. W. Lynn, C. J. Hood, and H. J. Kimble,
Phys. Rev. A {\bf 63}, 013401 (2000).

\bibitem{Knill01}
E. Knill, R. Laflamme, and G. J. Milburn, 
Nature {\bf 409}, 46 (2001).

\bibitem{Enk98}
S. J. Enk, J. I. Cirac, and P. Zoller,
Science {\bf 279}, 205 (1998).

\bibitem{Tregenna02}
B. Tregenna, A. Beige, and P. L. Knight,
Phys. Rev. A {\bf 65}, 032305 (2002).

\bibitem{Pan03}
J.-W. Pan, S. Gasparoni, R. Ursin, G. Weihs, and A. Zeilinger,
Nature {\bf 423}, 417 (2003).

\bibitem{Bennett00}
C. H. Bennett and D. P. DiVincenzo,
Nature {\bf 404}, 247 (2000).

\bibitem{Martini96}
F. De Martini, G. Di Giuseppe, and M. Marrocco,
Phys. Rev. Lett. {\bf 76}, 900 (1996).

\bibitem{Rauschenbeutel01}
A. Rauschenbeutel, P. Bertet, S. Osnaghi, G. Nogues, 
M. Brune, J. M. Raimond, and S. Haroche,
Phys. Rev. A {\bf 64}, 050301(R) (2001).

\bibitem{Browne03}
D. E. Browne and M. B. Plenio,
Phys. Rev. A {\bf 67}, 012325 (2003).

\bibitem{Domokos98}
P. Domokos, M. Brune, J. M. Raimond, and S. Haroche,
Eur. Phys. J. D {\bf 1}, 1 (1998).

\bibitem{Brattke01}
S. Brattke, B. T. H. Varcoe, and H. Walther,
Phys. Rev. Lett. {\bf 86}, 3534 (2001).

\bibitem{Solano03}
E. Solano, G. S. Agarwal, and H. Walther,
Phys. Rev. Lett. {\bf 90}, 027903 (2003).

\bibitem{Parkins95}
A. S. Parkins, P. Marte, P. Zoller, O. Carnal, and H. J. Kimble,
Phys. Rev. A {\bf 51}, 1578 (1995).

\bibitem{Hennrich00}
M. Hennrich, T. Legero, A. Kuhn, and G. Rempe
Phys. Rev. Lett. {\bf 85}, 4872 (2000).

\bibitem{Lange00}
W. Lange and H. J. Kimble,
Phys. Rev. A {\bf 61}, 063817 (2000).

\bibitem{Duan03}
L.-M. Duan, A. Kuzmich, and H. J. Kimble, 
Phys. Rev. A {\bf 67}, 032305 (2003).

\bibitem{Cirac97}
J. I. Cirac, P. Zoller, H. J. Kimble, and H. Mabuchi,
Phys. Rev. Lett. {\bf 78}, 3221 (1997).

\bibitem{Zippilli03}
S. Zippilli, D. Vitali, P. Tombesi, and J.-M. Raimond,
Phys. Rev. A {\bf 67}, 052101 (2003).

\bibitem{Kuhn02}
A. Kuhn, M. Hennrich, and G. Rempe,
Phys. Rev. Lett. {\bf 89}, 067901 (2002).

\bibitem{Rempe92}
G. Rempe, R. J. Thompson, and H. Kimble,
Opt. Lett. {\bf 17}, 363 (1992);
G. Rempe, private communication (2003).

\bibitem{Hood01}
C. J. Hood, H. J. Kimble, and Jun Ye,
Phys. Rev. A {\bf 64}, 033804 (2001).

\bibitem{Pelton02}
M. Pelton, C. Santori, J. Vu\v{c}kovi\'{c}, B. Zhang, G. S. Solomon, 
J. Plant, and Yoshihisa Yamamoto,
Phys. Rev. Lett. {\bf 89}, 233602 (2002).

\bibitem{Saavedra00}
C. Saavedra, K. M. Gheri, P. T\"orm\"a, J. I. Cirac, and P. Zoller,
Phys. Rev. A {\bf 61}, 062311 (2000).

\bibitem{Scheel01}
S. Scheel and D.-G. Welsch,
Phys. Rev. A {\bf 64},  063811 (2001).











\bibitem{Santos01}
M. Fran\c{c}a Santos, L. G. Lutterbach, S. M. Dutra, N. Zagury,
and L. Davidovich,
Phys. Rev. A {\bf 63}, 033813 (2001).

\bibitem{Gardiner85}
C. W. Gardiner and M. J. Collett,
Phys. Rev. A {\bf 31}, 3761 (1985).     

\bibitem{Knoell91}
L. Kn\"{o}ll, W. Vogel, and D.-G. Welsch,
Phys. Rev. A {\bf 43}, 543 (1991).      

\bibitem{Vogel94}
W. Vogel and D.-G. Welsch,
\textit{ Lectures on Quantum Optics},  Akademie Verlag GmbH, Berlin /
VCH Publishers, Inc., New York, (1994).

\bibitem{Walther03}
H. Walther, private communication (2003).

\bibitem{Khanbekyan03}
M. Khanbekyan, L. Kn\"{o}ll, and D.-G. Welsch,
Phys. Rev. A {\bf 67}, 063812 (2003).

\bibitem{Arfken}
G. Arfken, 
\textit{Mathematical Methods for Physicists}, Academic Press, Orlando,
(1985).

\bibitem{Ho01}
Ho Trung Dung, L. Kn\"{o}ll, and D.-G. Welsch,
Phys. Rev. A {\bf 64}, 013804 (2001).

\bibitem{Buck03}
J. R. Buck and H. J. Kimble,
Phys. Rev. A {\bf 67}, 033806 (2003).


\end{references}
\end{document}